\documentclass{webofc}
\usepackage[varg]{txfonts}
\usepackage{graphicx,amsmath,wrapfig,epsfig}
\woctitle{6th International Conference on Physics Opportunities 
at an ElecTron-Ion Collider}
\begin{document}
\title{Progress on nuclear modifications of structure functions}
\author{S. Kumano\inst{1,2}} 
\institute{
KEK Theory Center, Institute of Particle and Nuclear Studies,
KEK, \\
\ \, 1-1, Ooho, Tsukuba, Ibaraki, 305-0801, Japan
\and
J-PARC Branch, KEK Theory Center, Institute of
Particle and Nuclear Studies, KEK, \\
\ \, and Theory Group, Particle and Nuclear Physics Division, 
  J-PARC Center, \\
\ \, 203-1, Shirakata, Tokai, Ibaraki, 319-1106, Japan}

\abstract{
We report progress on nuclear structure functions,
especially on their nuclear modifications and a new tensor structure function
for the deuteron. To understand nuclear structure functions
is an important step toward describing nuclei and QCD matters from
low to high densities and from low to high energies in terms of 
fundamental quark and gluon degrees of freedom beyond 
conventional hadron and nuclear physics. It is also practically important
for understanding new phenomena in high-energy heavy-ion collisions at 
RHIC and LHC. 
Furthermore, since systematic errors of current neutrino-oscillation
experiments are dominated by uncertainties of neutrino-nucleus interactions,
such studies are valuable for finding new physics beyond current framework. 
Next, a new tensor-polarized structure function $b_1$
is discussed for the deuteron. There was a measurement by HERMES;
however, its data are inconsistent with the conventional
convolution estimate based on the standard deuteron model
with D-state admixture. This fact suggests that a new hadronic
phenomenon should exist in the tensor-polarized deuteron at high energies,
and it will be experimentally investigated at JLab from the end of 2010's.
}

\maketitle

\section{Introduction}
\label{intro}

Nuclear structure functions are different from corresponding ones
of the nucleon. Such nuclear modifications have been measured
from relatively small $x$ ($\sim 0.005$) to large $x$ ($\sim 0.7$)
mainly in charged-lepton deep inelastic scattering (DIS). 
Physics mechanisms for the nuclear effects are different 
in each $x$ region \cite{nuclear-summary}. 
At small $x$, a virtual photon becomes $q\bar q$ (or vector mesons), 
which interacts with a nucleus by strong interactions. It leads to 
nuclear shadowing phenomena due to multiple interactions in a nucleus.
At medium $x$, there are negative contributions from nuclear binding 
and possible internal nucleon modifications. The modifications are
positive at large $x$ due to nucleon's Fermi motion in a nucleus.
Although the major nuclear-modification mechanisms are known,
we cannot calculate the structure functions precisely,
typically less than 10\% accuracy, by the theoretical models.
For actual application to heavy-ion physics and neutrino
reactions, it is practically not appropriate to reply on such models.
For example, neutrino-nucleus cross sections need to be
calculated within 5\% accuracy in future leptonic CP
violation measurements \cite{J-PARC-th-neutrino-A}. 
Therefore, it is desirable to use global analysis results as a model 
for the nuclear parton distribution functions (NPDFs)
\cite{hkn,other-npdfs}, in the same way as the global analysis PDFs 
are used for calculating cross sections at LHC to find
physics beyond the standard model.

The NPDFs are determined by analyzing world data on high-energy
nuclear reactions, including charged-lepton DIS, neutrino DIS, 
Drell-Yan, and hadron productions.
Currently, there are two major issues in the NPDF studies.
One is that gluon shadowing is not determined from the current
measurements \cite{hkn,other-npdfs}, and the other is that 
modifications suggested by the neutrino DIS data could be different 
from the ones which are inferred from the charged-lepton DIS
\cite{nCTEQ-nu}. On the first point, the issue is due partly to the fact
that a lepton-nucleus collider similar to HERA does
not exist to observe scaling violation at small $x$.
However, the situation could improve because LHC started
producing data, which are sensitive to small-$x$ physics.
On the second point, somewhat conflicting results are
obtained among different analysis groups. It is desirable 
to clarify the situation by another serious analysis
to discuss the details of analysis conditions and data handling.
In this report, we discuss the situation of these studies
in Sec.\,\ref{npdfs}.

As the second topic, we explain tensor structure functions
of the deuteron in charged-lepton DIS with a polarized deuteron
\cite{sk-b1-summary}.
The deuteron structure is well known at low energies and it 
is described by a proton-neutron S-wave bound state 
with small D-wave admixture. Since the deuteron is a spin-1 hadron,
it has new polarized structure functions, $b_1$, $b_2$, $b_3$,
and $b_4$ in addition to the four structure functions,
$F_1$, $F_2$, $g_1$, and $g_2$,
which exist for the spin-1/2 nucleon in the charged-lepton DIS.
The structure functions $b_1$ and $b_2$ are leading-twist ones, 
which are related by the Callan-Gross like relation $2x b_1 =b_2$
in the scaling limit. The $b_3$ and $b_4$ are higher-twist ones.

There are only a few $b_1$ data measured by the HERMES collaboration
\cite{hermes05}. Although much accurate measurements are needed, 
there is already an interesting hint in the data toward a new discovery.
The most conventional way to estimate $b_1$ theoretically is 
to use a convolution model, where $b_1$ is calculated
by an unpolarized structure function of the nucleon
with nucleon-momentum distributions including the D wave.
The HERMES data are an order-of-magnitude larger than
this standard model predication. It indicates that $b_1$ cannot be
understood by the conventional model, and a new hadronic physics
should be introduced for interpretation of the HERMES data.
Fortunately, an experimental proposal to JLab was approved for $b_1$,
and its measurement will start around the year of 2019 \cite{Jlab-b1}.
It is a good opportunity to find a new exotic hadronic phenomenon
in the simplest nucleus, deuteron. The deuteron tensor structure
used to play an important role in low-energy nuclear physics.
It is now time to understand the tensor structure in terms of
quark and gluon degrees of freedom with new hadron physics.
We explain this current situation as the second topic in Sec.\,\ref{b1}.

\section{Nuclear parton distribution functions}
\label{npdfs}

Nuclear parton distributions are modified from the ones for the nucleon.
The space is limited for this article, so that we do not address ourselves
to physics mechanisms. An interested reader may read summary articles
in Ref.\,\cite{nuclear-summary}. We focus our discussions on the status
of NPDFs and an application to neutrino physics.
The NPDFs are determined by a global analysis of world data on high-energy
lepton-nucleus, proton-nucleus, and nucleus-nucleus reactions.
The distributions are parametrized, and optimum parameters are determined
by a $\chi^2$ analysis. There are a few groups which investigate
the NPDFs. There are two types for the parametrization.
One is to obtain modifications from typical nucleonic PDFs,
and the other is to obtain the NPDFs directly. The latter one 
is a general way of analysis on the same footing with the nucleonic
PDF analysis, whereas it is easier to impose physical constrains
in the former one because the determined NPDFs could become
unphysical distributions due to the lack of data especially at 
extreme kinematical conditions. If we were to have sufficient experimental
data, both results should agree with each other.

There are two major issues in the determination of NPDFs. First, nuclear
modifications of the gluon distribution cannot be fixed due to the lack of
data which are sensitive to them. Second, some inconsistencies were
pointed out for the modifications between charged-lepton and neutrino
scattering processes \cite{nCTEQ-nu}, although some other analysis do not
show such inconsistency clearly. In the following, we first explain 
a connection to neutrino physics and it is partially related 
to the second issue.

\begin{wrapfigure}[15]{r}{0.38\textwidth}
\vspace{-0.5cm}
   \begin{center}
       \epsfig{file=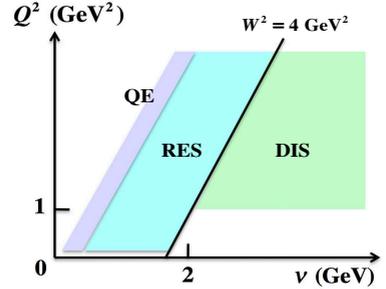,width=0.36\textwidth} \\
   \end{center}
\vspace{-0.70cm}
\caption{
  Kinematical regions of neutrino-nucleus scattering. Here,
  QE, RES, and DIS indicate quasi-elastic, resonance, and deep inelastic
  scattering.
}
\label{fig:neutirno-kinematics}
\end{wrapfigure}

Precise NPDFs are needed for finding any new physics in high-energy
nuclear reactions such as properties of quark-gluon matters in 
heavy-ion collisions at RHIC and LHC. There is another important
application to neutrino physics. The statistical errors in
neutrino-oscillation measurements become smaller and smaller, 
but their systematic errors are still dominated by neutrino-nucleus interactions,
which is an obstacle for a new discovery.
There are three major kinematical regions as shown in 
Fig. \ref{fig:neutirno-kinematics}:
quasi-elastic (QE), resonance (RES), and deep inelastic scattering (DIS).
There are not definite boundaries in the sense that $W^2$ and $Q^2$ cut
values depend on researchers.
However, a usual choice is the cut $W^2 \ge 4$ GeV$^2$ for the DIS,
and $Q^2$ should be large enough, typically $Q^2 \ge 1$ GeV$^2$.
These three regions have been investigated by different types of physicists.
In the current neutrino experiments, except for the ultra-high energy
IceCube experiment, typical energies are from several hundred MeV
to a few dozen GeV. Therefore, all of these kinematical cross sections
should be precisely known for reducing the systematical uncertainties
in neutrino oscillation measurements. For future leptonic CP violation
measurements, the cross sections need to be understood
within the 5\% level. A project is in progress to provide a code to calculate
the cross section in any kinematical range by combining theories
of different regions at the J-PARC branch of the KEK theory center
\cite{J-PARC-th-neutrino-A}.
In the following, we discuss only the DIS studies and such a unification
effort will be explained elsewhere.

If there were no nuclear modification, the NPDFs are simple addition
of proton and neutron contributions. However, from experimental measurements,
nuclear medium effects are typically 10-20\% depending on
a nuclear size. Such medium effects are parametrized
by the function $w_i$ in Ref.\,\cite{hkn} as 
\begin{equation}
f_i^A (x,Q_0^2) = w_i (x,A,Z) \, 
    \frac{1}{A} \left[ Z\,f_i^{p} (x,Q_0^2) 
                + N f_i^{n} (x,Q_0^2) \right] ,
\label{eqn:fia}
\end{equation}
where $p$ and $n$ indicate the proton and neutron,
$A$, $Z$, and $N$ are mass number, atomic number, and neutron number,
$i$ indicates a type of distribution ($i=u_v,\ d_v,\ \bar q,\ g$),
and $Q_0^2$ is the initial $Q^2$ scale. 
The functions $w_i$ are expressed by a number of parameters,
which are then determined by a global analysis.
The kinematical range of $x$ is $0<x<A$ for a nucleus.
Therefore, the functional form of Eq.\,(\ref{eqn:fia}) cannot
describe the region $x>1$. However, there is no DIS data in
such a large-$x$ region, so it is not a serious problem at this stage.
In the HKN analysis \cite{hkn}, the nuclear modification functions
are expressed in terms of the parameters,
$\alpha$, $\beta$, $a_i$, $b_i$, $c_i$, and $d_i$ as
\begin{equation}
w_i (x,A,Z) = 1 + \left( {1 - \frac{1}{{A^{\alpha} }}} \right)
\frac{{a_i  + b_i x + c_i x^2  + d_i x^3 }}{{(1 - x)^{\beta} }} ,
\\
\end{equation} 
at $Q_0^2=1 \ {\rm GeV}^2$. The parameters are determined
by a global $\chi^2$ analysis with the standard DGLAP evolution
equation in comparison with the world data.

\vfill\eject

\begin{wrapfigure}[12]{r}{0.40\textwidth}
\vspace{+0.2cm}
\hspace{-0.16cm} 
       \epsfig{file=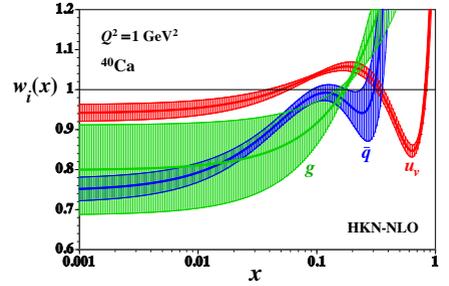,width=0.40\textwidth} 
\vspace{-0.35cm}
\caption{HKN nuclear modifications for $^{40}$Ca at $Q^2=1$ GeV$^2$.}
\label{fig:nuclear-pdfs}
\end{wrapfigure}

As an example, nuclear modifications of $^{40}$Ca PDFs
and their uncertainties are shown 
for the next-to-leading (NLO) order
HKN parametrization \cite{hkn} in Fig. \ref{fig:nuclear-pdfs}
at $Q^2=1$ GeV$^2$. There are some differences in
the NPDFs from other groups \cite{other-npdfs} at large $x$ 
especially for the antiquark and gluon distributions 
because there are few data which constrain them.
In addition, large differences exist in the gluon distributions
at small $x$. The nucleonic gluon distribution is constrained
by the scaling violation of $F_2$ from the HERA $ep$-collider
experiments, whereas there is no such a measurement for nuclei at small $x$.
However, the situation may improve in the near future
due to LHC measurements with heavy ions. Using the obtained NPDFs
together with DGLAP evolution equations,
one can calculate nuclear modifications
of structure functions in neutrino scattering.
Some inconsistencies were reported between the charged-lepton DIS
and neutrino DIS structure functions in Ref.\,\cite{nCTEQ-nu};
however, the differences are not clear in other groups
\cite{other-npdfs}. Our analysis is now in progress by including
neutrino data for clarifying this issue as an independent trial.

\section{New spin structure function \boldmath{$b_1$} for deuteron}
\label{b1}

There are four new structure functions, $b_1$, $b_2$, $b_3$, and $b_4$,
for the spin-1 hadron in charged-lepton deep inelastic scattering. 
They are defined in the hadron tensor as \cite{sk-b1-summary}:
\vspace{-0.10cm}
\begin{align}
W_{\mu \nu}^{\,\lambda_f \lambda_i}
   = & -F_1 \, \hat{g}_{\mu \nu} 
     +\frac{F_2}{M \nu} \hat{p}_\mu \hat{p}_\nu 
     + \frac{ig_1}{\nu}\epsilon_{\,\mu \nu \lambda \sigma} \,
        q^{\,\lambda} s^{\,\sigma}  
     +\frac{i g_2}{M \nu ^2}\epsilon_{\,\mu \nu \lambda \sigma} \,
      q^{\,\lambda} \, (p \cdot q s^{\,\sigma} - s \cdot q p^{\,\sigma} )
\notag \\
& 
     -b_1 r_{\mu \nu} 
     + \frac{1}{6} b_2 (s_{\mu \nu} +t_{\mu \nu} +u_{\mu \nu}) 
     + \frac{1}{2} b_3 (s_{\mu \nu} -u_{\mu \nu}) 
     + \frac{1}{2} b_4 (s_{\mu \nu} -t_{\mu \nu}) ,
\label{eqn:w-1}
\\[-0.70cm] \nonumber
\end{align}
where $p$ and $q$ are spin-1 hadron and virtual-photon momenta,
$\nu$ is the energy transfer, and
$r_{\mu \nu}$, $s_{\mu \nu}$, $t_{\mu \nu}$, and $u_{\mu \nu}$
are defined by the polarization vector of the spin-one hadron $E^\mu$
in Ref.\,\cite{sk-b1-summary}.
The spin vector $s$ is expressed by the polarization vector as
$ (s_{\lambda_f \lambda_i})^{\,\mu}
      = - i \epsilon ^{\,\mu \nu \alpha \beta} 
                E^*_\nu (\lambda_f) E_\alpha (\lambda_i) p_\beta / M $
where the initial and final polarization vectors are denoted as
$E^{\,\mu} (\lambda_i)$ and $E^{\,\mu} (\lambda_f)$
with the spin states $\lambda_i$ and $\lambda_f$,
and $M$ is the mass of the spin-1 hadron.
The coefficients of $b_1$, $b_2$, $b_3$ and $b_4$ are defined as
symmetric under $\mu \leftrightarrow \nu$ in Eq.\,(\ref{eqn:w-1}), 
and they vanish under the spin average. 
They are also defined so that $b_1$ and $b_2$ are twist-two
functions to satisfy the Callan-Gross type relation,
$2x b_1 =b_2$. The $b_3$ and $b_4$ are higher-twist structure functions.
The twist-two structure function $b_1$ is expressed
in terms of the tensor-polarized parton distributions 
$\delta_{_T} q$ and $\delta_{_T} \bar q$ as
\vspace{-0.30cm}
\begin{align}
b_1 (x,Q^2) = \frac{1}{2} \sum_i e_i^2 
      \, \left [ \delta_{_T} q_i (x,Q^2) 
      + \delta_{_T} \bar q_i (x,Q^2)   \right ] , \ \ \
       \delta_{_T} q_i  \equiv q_i^0 - \frac{q_i^{+1}+q_i^{-1}}{2} .
\label{eqn:b1-parton}
\\[-0.80cm] \nonumber
\end{align}
Here, $i$ indicates the flavor of a quark, $e_i$ is a quark charge,
and $q_i^\lambda$ indicates an unpolarized-quark distribution
in the hadron spin state $\lambda$. 
The tensor-polarized distribution is much different from 
the longitudinally-polarized one 
$\Delta q = q_\uparrow - q_\downarrow$, where $\uparrow$ and $\downarrow$
indicate the polarization of a quark along the hadron-spin direction,
in the sense that it is ``unpolarized"-quark distribution
in a tensor-polarized spin-1 hadron.

The structure function $b_1$ can be calculated by the standard
convolution picture for the deuteron in terms of parton-momentum
distributions convoluted with nucleon momentum distributions.
The $b_1$ is associated with tensor structure of the deuteron,
so that D-wave admixture should be properly considered
for the convolution estimation. However, it is surprising
to find an order-of-magnitude difference between such
a standard-model estimate and experimental data obtained
by the HERMES collaboration, although the data have large errors.
There are other theoretical-model calculations;
however, we do not step into such theoretical models in this
article, and we discuss probable tensor-polarized distributions
from the HERMES data.

Such a parametrization was studied in Ref.\,\cite{sk-tensor-pdf}.
We consider the following constraint in the form of $b_1$ sum rule
\cite{ck-sum}, which was obtained in a parton model and it is
similar to the Gottfried sum:
\vspace{-0.20cm}
\begin{alignat}{2}
\int dx \, b_1 (x) 
   & = 0 &
   & + \frac{1}{9} \int dx
      \, \left [ \, 4 \, \delta_{_T} \bar u (x) +  \delta_{_T} \bar d (x) 
                     +   \delta_{_T} \bar s (x)  \, \right ] ,
\nonumber \\
\int \frac{dx}{x} \, [F_2^p (x) - F_2^n (x) ]
     & =  \frac{1}{3} &
 & +\frac{2}{3} \int dx \, [ \bar u(x) - \bar d(x) ] . 
\label{eqn:b1-sum-gottfried}
\\[-0.70cm] \nonumber
\end{alignat}
As the Gottfried-sum-rule violation created a field of
flavor-symmetric light-antiquark distributions ($\bar u \ne \bar d$)
\cite{ubar-dbar}, a finite $b_1$ sum could indicate a finite tensor-polarized 
antiquark distribution, which cannot be expected from ordinary 
theoretical ideas. For the time being as the first-step study, 
we neglect such exotic contributions and consider the condition 
$\int dx b_1 (x)=0$ for constraining the tensor distributions.

As the first trial, we consider the case that a certain fraction
($\delta_{_T} w$) of unpolarized PDFs is tensor polarized 
in the deuteron \cite{sk-tensor-pdf}:
$\delta_{_T} q_{iv}^D (x) = \delta_{_T} w(x) \, q_{iv}^D (x)$, 
$\delta_{_T} \bar q_i^D (x) 
           = \alpha_{\bar q} \, \delta_{_T} w(x) \, \bar q_i^D (x)$
for the tensor-polarized parton distributions.
Here, we assume a common function $\delta _{_T} w (x)$
for both quarks and antiquarks except for a different 
overall constant $\alpha_{\bar q}$.
Now, the problem becomes how to determine $\delta_{_T} w (x)$.
We assume that the valence-quark distributions 
should satisfy the sum $\int dx (b_1)_{\text{valence}}=0$.
Then, the function $\delta_{_T} w(x)$ should have a node at least, 
so we may take the parametrization
$ \delta_{_T} w(x) = a x^b (1-x)^c (x_0-x) $.
Here, $a$, $b$, and $c$ are the parameters determined by the analysis,
and the node position $x_0$ is expressed by the other parameters as
$ x_0= \int dx \, x^{\, b+1} (1-x)^c ( u_v + d_v )  /
       \int dx \, x^{\, b}   (1-x)^c ( u_v + d_v ) $
due to the sum rule. Such a node also exists in $b_1$
calculated by the convolution model with the D-state admixture. 

\vspace{-0.0cm}
\noindent
\begin{figure}[b!]
\vspace{-0.5cm}
\parbox[t]{0.48\textwidth}{
   \begin{center}
    \hspace{-1.2cm}
    \epsfig{file=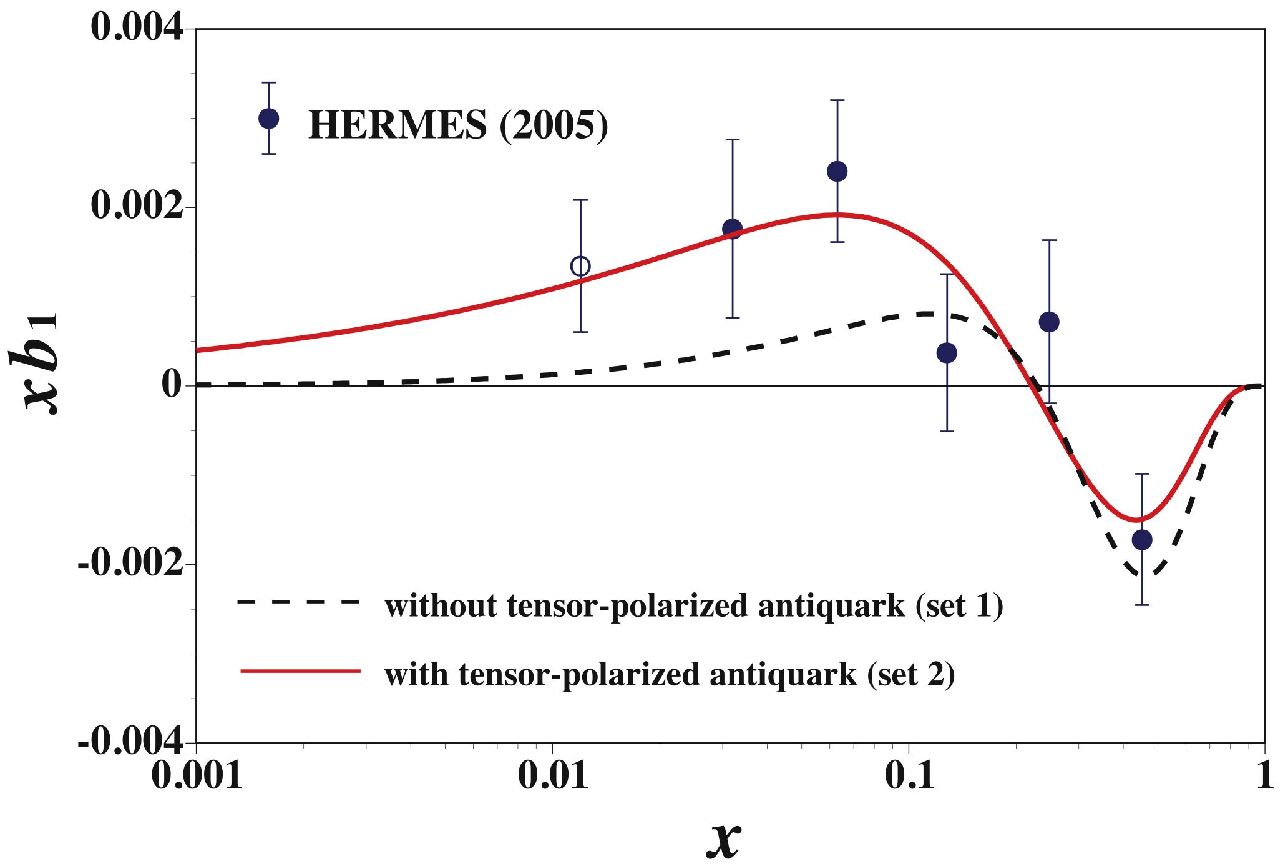,width=0.38\textwidth}
   \end{center}
\vspace{-0.7cm}
\caption{Obtained $b_1$ and HERMES data \cite{sk-tensor-pdf}.}
\label{fig:b1-hermes-data}
}\hfill
\parbox[t]{0.50\textwidth}{
   \begin{center}
    \epsfig{file=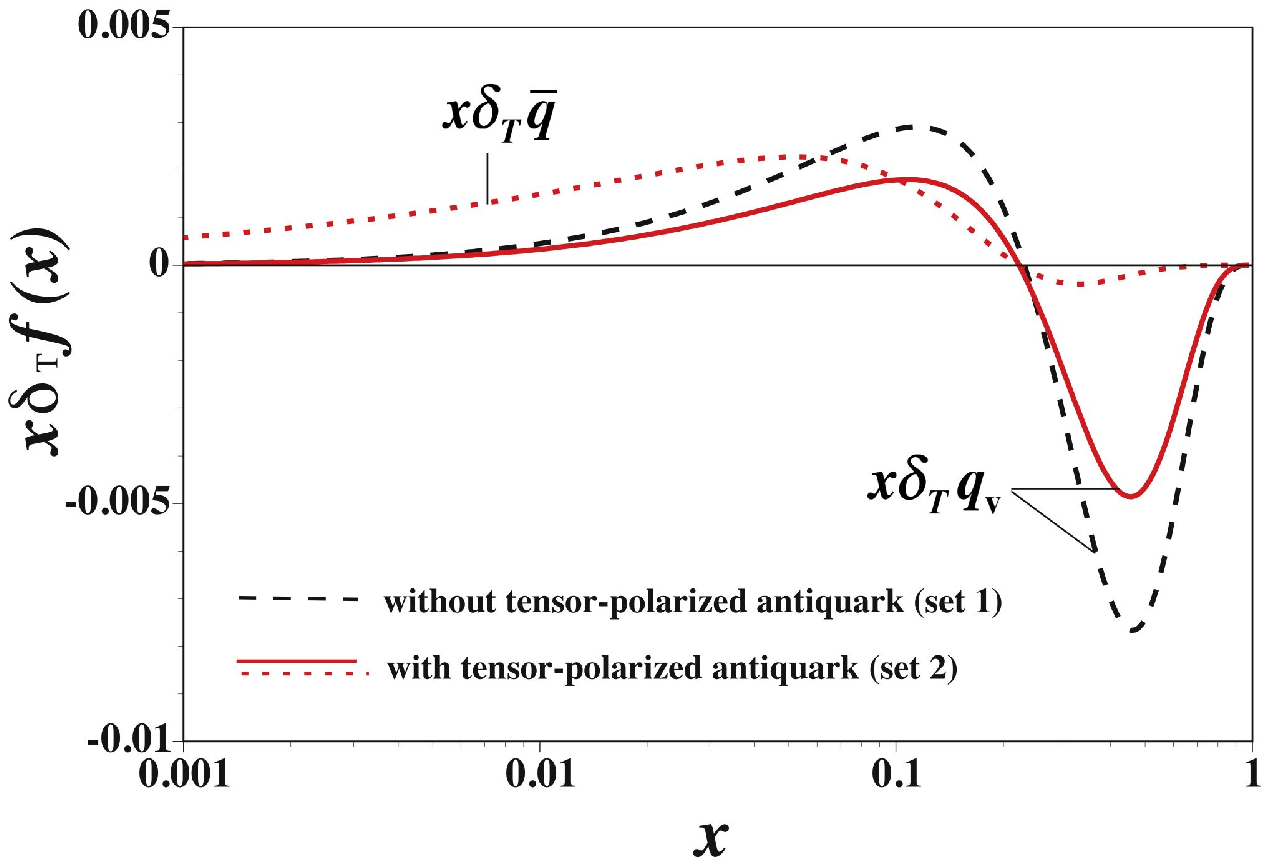,width=0.38\textwidth}
   \end{center}
\vspace{-0.7cm}
\caption{Determined tensor-polarized distributions \cite{sk-tensor-pdf}.}
\label{fig:tensor-distribution}
}
\end{figure}
\vspace{-0.35cm}

We tried two scenarios depending on whether or not the antiquark 
tensor polarization exists:
\vspace{-0.10cm}
\begin{itemize}
\item Set 1: Tensor-polarized antiquark distributions are terminated
             ($\alpha_{\bar q} = 0$).
\vspace{-0.10cm}
\item Set 2: Finite tensor-polarized antiquark distributions
             are allowed ($\alpha_{\bar q}$ is a parameter).
\end{itemize}
The parameters are determined by a $\chi^2$ analysis in each case.
Determined $b_1$ and tensor-polarized PDFs are shown in Figs.
\ref{fig:b1-hermes-data} and \ref{fig:tensor-distribution}.
There are two curves in Fig. \ref{fig:b1-hermes-data} for
the sets 1 and 2. It is obvious from this figure,
within the restriction of the used parametrization,
the HERMES data cannot be explained without the antiquark
tensor polarization. Namely, $\chi^2$ is much larger for the set-1 analysis.
The obtained tensor distributions are shown in 
Fig. \ref{fig:tensor-distribution}.
The dashed curve indicates the valence-quark distribution in the set 1.
The solid and dotted curves indicate the valence-quark
and antiquark distributions in the set 2. 
The valence-quark distribution is negative at large $x$ ($>0.3$) 
and then it turns into positive at $x \simeq 0.2$.
The antiquark distributions are especially large at small $x$ ($<0.1$).
Since there are only a few data points 
at this stage, there are large uncertainties for the distributions.

In future, we need theoretical and experimental efforts on $b_1$ physics.
On the theory side, we need to understand the obtained tensor-polarized
distributions in Fig. \ref{fig:tensor-distribution}
by appropriate theoretical models. The standard deuteron model
with the D-state admixture cannot explain the distributions
because a typical convolution-model estimate is one-order-of-magnitude
smaller than the values in Fig. \ref{fig:tensor-distribution}. 
We need to investigate more exotic hadronic mechanisms.
At the same time, experimental efforts are needed to measure
$b_1$ much accurately because there are only a few data points
as shown in Fig. \ref{fig:b1-hermes-data} with the large errors.
However, it is fortunate that the JLab $b_1$ experiment was
approved, and the measurement will start in a few years \cite{Jlab-b1}.
On the other hand, there are possibilities for investigating
the antiquark tensor polarization by polarized Drell-Yan processes,
such as $p +\vec d \to \mu^+ \mu^- +X$ and $\pi +\vec d \to \mu^+ \mu^- +X$,
at hadron facilities, such as Fermilab-MI, RHIC, CERN-COMPASS, J-PARC, 
GSI-FAIR, and U70 \cite{tensor-dy}. For example, a tensor spin symmetry 
in $p \vec d$ Drell-Yan is given by
\vspace{-0.10cm}
\begin{equation}
A_{UQ_0} \text{(large $x_F$)}\simeq
           \frac{\sum_i e_i^2 \, \bar q_i (x_1) \, \delta_{_T} q_i (x_2) }
                {\sum_i e_i^2 \, \bar q_i (x_1) \, q_i (x_2) } ,
\vspace{-0.10cm}
\end{equation}
so that $\delta_{_T} q_i$ can be measured directly.
It is similar to the case that $\bar u/\bar d \ne 1$ was
clarified by the Drell-Yan experiment although it was suggested
by the Gottfried-sum-rule violation \cite{ubar-dbar}.

\vspace{-0.2cm}
\section*{Acknowledgements}
\vspace{-0.1cm}
This work was supported by Ministry of Education, Culture, Sports, 
Science and Technology (MEXT) KAKENHI Grant No. 25105010.

\vspace{-0.2cm}


\end{document}